# Scientific reticence and sea level rise

J E Hansen

NASA Goddard Institute for Space Studies, 2880 Broadway, New York, NY 10025, USA

E-mail: jhansen@giss.nasa.gov

**Abstract**
I suggest that a 'scientific reticence' is inhibiting communication of a threat of potentially large sea level rise.  Delay is dangerous because of system inertias that could create a situation with future sea level changes out of our control.  I argue for calling together a panel of scientific leaders to hear evidence and issue a prompt plain-written report on current understanding of the sea level change issue.

Scientific reticence and sea level rise

**1. Introduction**

  I suggest that 'scientific reticence', in some cases, hinders communication with the public about dangers of global warming.  If I am right, it is important that policy-makers recognize the potential influence of this phenomenon.
  Scientific reticence may be a consequence of the scientific method.  Success in science depends on objective skepticism.  Caution, if not reticence, has its merits.  However, in a case such as ice sheet instability and sea level rise, there is a danger in excessive caution.  We may rue reticence, if it serves to lock in future disasters.
  Barber (1961) describes a 'resistance by scientists to scientific discovery', with a scholarly discussion of several sources of cultural resistance.  There are aspects of the phenomenon that Barber discusses in the 'scientific reticence' that I describe, but additional factors come into play in the case of global climate change and sea level rise.
  I can illustrate 'scientific reticence' best via personal experiences.  The examples are relevant to the Intergovernmental Panel on Climate Change (IPCC) process of consensus building, specifically to the issue of possible sea level rise.

**2. The Court Case.**

  'Scientific reticence' leapt to mind as I was being questioned, and boxed-in, by a lawyer for the plaintiff in Automobile Manufacturers versus California Air Resources Board (Auto Manufacturers 2006).  I conceded that I was not a glaciologist.  The lawyer then, with aplomb, requested that I identify glaciologists who agreed publicly with my assertion that sea level was likely to rise more than one meter this century if greenhouse gas emissions followed an IPCC business-as-usual (BAU) scenario: "Name one!"
  I could not, instantly.  I was dismayed, because, in conversation and e-mail exchange with relevant scientists I sensed a deep concern about likely consequences of BAU global warming for ice sheet stability.  What would be the legal standing of such a lame response as 'scientific reticence'?  Why would scientists be reticent to express concerns about something so important?
  I suspect the existence of what I call the "John Mercer effect".  Mercer (1978) suggested that global warming from burning of fossil fuels could lead to disastrous disintegration of the West Antarctic ice sheet, with sea level rise of several meters worldwide.  This was during the era when global warming was beginning to get attention from the United States Department of Energy and other science agencies.  I noticed that scientists who disputed Mercer, suggesting that his paper was alarmist, were treated as being more authoritative.
  It was not obvious who was right on the science, but it seemed to me, and I believe to most scientists, that the scientists preaching caution and downplaying the dangers of climate change fared better in receipt of research funding.  Drawing attention to the dangers of global warming may or may not have helped increase funding for relevant scientific areas, but it surely did not help individuals like Mercer who stuck their heads out.  I could vouch for that from my own experience.  After I published a paper (Hansen *et al* 1981) that described likely climate effects of fossil fuel use, the Department of Energy reversed a decision to fund our research, specifically highlighting and criticizing aspects of that paper at a workshop in Coolfont, West Virginia and in publication (MacCracken 1983).
  I believe there is a pressure on scientists to be conservative.  Papers are accepted for publication more readily if they do not push too far and are larded with caveats.  Caveats are essential to science, being born in skepticism, which is essential to the process of investigation and verification.  But there is a question of degree.  A tendency for 'gradualism' as new evidence comes to light may be ill-suited for communication, when an issue with short time fuse is concerned.
  However, these matters are subjective.  I could not see how to prove the existence of a 'scientific reticence' about ice sheets and sea level.  Score one for the plaintiff, and their ally and 'friend of the court', the United States federal government.

Scientific reticence and sea level rise

**3. On the Ice**

A field glaciologist, referring to a moulin on Greenland, said: "the whole damned ice sheet is going to go down that hole!" He was talking about his expectations, under the assumption of continued unchecked growth of global GHG emissions. Field glaciologists have been doing a good job of reporting current trends on the ice sheets. It is translation of field data into conclusions needed by the public and policymakers that is at issue.

Ice sheet disintegration, unlike ice sheet growth, is a wet process that can proceed rapidly. Multiple positive feedbacks accelerate the process once it is underway. These feedbacks occur on and under the ice sheets and in the nearby oceans.

A key feedback on the ice sheets is the 'albedo flip' (Hansen *et al* 2007) that occurs when snow and ice begin to melt. Snow-covered ice reflects back to space most of the sunlight striking it. However, as warming causes melting on the surface, the darker wet ice absorbs much more solar energy. Most of resulting melt water burrows through the ice sheet, lubricates its base, and thus speeds discharge of icebergs to the ocean (Zwally *et al.* 2002).

The area with summer melt on Greenland increased from ~450,000 km$^2$ when satellite observations began in 1979 to more than 600,000 km$^2$ in 2002 (Steffen *et al* 2004). Linear fit to data for 1992-2005 yields an increase of melt area of +40,000 km$^2$ per year (Tedesco 2007), but this rate may be exaggerated by the effect of stratospheric aerosols from the 1991 volcanic eruption of Mount Pinatubo, which reduced summer melt in 1992. Summer melt on West Antarctica has received less attention than on Greenland, but it is more important. Satellite QuickSCAT radiometer observations reveal increasing areas of summer melt on West Antarctica and an increasing melt season length during the period 1999-2005 (Nghiem *et al* 2007).

The key role of the ocean, in the matter of ice sheet stability, is as a conduit for excess global-scale heating that eventually leads to melting of ice. The process begins with increasing human-made greenhouse gases, which cause the atmosphere to be more opaque at infrared wavelengths. The increased atmospheric opacity causes heat radiation to space to emerge from a higher level, where it is colder, thus decreasing radiation of heat to space. As a result, the Earth is now out of energy balance by between 0.5 and 1 W/m$^2$ (Hansen *et al* 2005).

This planetary energy imbalance is itself now sufficient to melt ice corresponding to one meter of sea level rise per decade, if the energy were used entirely for that purpose (Hansen *et al* 2005). However, so far most of the excess energy has been going into the ocean. Acceleration of ice sheet disintegration requires tapping into ocean heat, which occurs primarily in two ways (Hansen 2005): (1) increased velocity of outlet glaciers (flowing in rock-walled channels) ice streams (bordered mainly by slower moving ice), and thus increased flux and subsequent melting of icebergs discharged to the open ocean, and (2) direct contact of ocean and ice sheet (underneath and against fringing ice shelves). Ice loss from the second process has a positive feedback on the first process: as buttressing ice shelves melt, ice stream velocity increases.

Positive feedback from loss of buttressing ice shelves is relevant to some Greenland ice streams, but the West Antarctic ice sheet, which rests on bedrock well below sea level (Thomas *et al* 2004), will be affected much more. Loss of ice shelves provides exit routes with reduced resistance for ice from further inland, as suggested by Mercer (1978) and earlier by Hughes (1972). Warming ocean waters are now thinning some West Antarctic ice shelves by several meters per year (Payne *et al* 2004; Shepherd *et al* 2004).

The Antarctic Peninsula recently provided a laboratory to study feedback interactions, albeit for ice volumes less than those in the major ice sheets. Combined actions of surface melt (van den Broeke 2005) and ice shelf thinning from below (Shepherd *et al* 2003) led to sudden collapse of the Larsen B ice shelf, which was followed by acceleration of glacial tributaries far inland (Rignot *et al* 2004; Scambos *et al.* 2004). The summer warming and melt that preceded ice shelf collapse (Fahnestock *et al.* 2002; Vaughan *et al* 2003) was no more than the global warming expected this century under BAU scenarios, and only a fraction of expected West Antarctic warming with realistic polar amplification of global warming.

Modeling studies yield increased ocean heat uptake around West Antarctica and Greenland due to increasing human-made greenhouse gases (Hansen *et al* 2006b). Observations show a warming ocean around West Antarctica (Shepherd *et al* 2004), ice shelves thinning several meters per year (Rignot and Jacobs 2002; Payne *et al* 2004), and increased iceberg discharge (Thomas *et al* 2004). As discharge of ice increases from a disintegrating ice sheet, as occurs with all deglaciations, regional cooling by the icebergs is significant, providing a temporary negative feedback (Hansen 2005). However, this cooling effect is limited on global scale as shown by comparison with the planetary energy imbalance, which is sufficient to melt ice equivalent to about one meter of sea level per decade (Table S1 of Hansen et al. 2005). Indeed, cooling of the ocean surface by melting ice increases the planetary energy imbalance, thus supplying additional energy for ice melt, so the planetary energy imbalance should not be thought of as a limit on the rate of ice melt.

Global warming should also increase snowfall accumulation rates in ice sheet interiors because of the higher moisture content of the warming atmosphere. Despite high variability on interannual and decadal time scales, and limited Antarctic warming to date, observations tend to support this expectation for both Greenland and Antarctica (Rignot and Thomas 2002; Johannessen *et al* 2005; Davis *et al* 2005; Monaghan *et al* 2006). Indeed, some models (Wild *et al* 2003) have ice sheets growing overall with global warming, but those models do not include realistic processes of ice sheet disintegration. Extensive paleoclimate data confirm the common sense expectation that the net effect is for ice sheets to shrink as the world warms.

The most compelling data for the net change of ice sheets is provided by the gravity satellite mission GRACE, which shows that both Greenland (Chen *et al* 2006) and Antarctica (Velicogna *et al* 2006) are losing mass at substantial rates. The most recent analyses of the satellite data (S. Klosco *et al* priv. comm.) confirm that Greenland and Antarctica are each losing mass at a rate of about 150 cubic kilometers per year, with the Antarctic mass loss primarily in West Antarctica. These rates of mass loss are at least a doubling of rates of several years earlier, and only a decade earlier these ice sheets were much closer to mass balance (Casenave 2006).

The Antarctic data are the most disconcerting. Warming there has been limited in recent decades, at least in part due to effects of ozone depletion (Shindell and Schmidt 2004). The fact that West Antarctica is losing mass at a significant rate suggests that the thinning ice shelves are already beginning to have an effect on ice discharge rates. Warming of the ocean surface around Antarctica (Hansen *et al* 2006a) is small compared with the rest of world, consistent with climate model simulations (IPCC 2007), but that limited warming is expected to increase (Hansen *et al* 2006b). The detection of recent, increasing summer surface melt on West Antarctica (Nghiem *et al* 2007) raises the danger that feedbacks among these processes could lead to nonlinear growth of ice discharge from Antarctica.

**4. Urgency: This Problem is Non-Linear!**

IPCC business-as-usual (BAU) scenarios are constructs in which it is assumed that emissions of $CO_2$ and other greenhouse gases will continue to increase year after year. Some energy analysts take it as almost a law of physics that such growth of emissions will continue in the future. Clearly, there is not sufficiently widespread appreciation of the implications of putting back into the air a large fraction of the carbon stored in the ground over epochs of geologic time. Climate forcing due to these greenhouse gases would dwarf the climate forcing for any time in the past several hundred thousand years, when accurate records of atmospheric composition are available from ice cores.

However, the long-term global cooling and increase of global ice through the Plio-Pleistocene provides an even more poignant illustration of the implications of continued BAU burning of fossil fuels. The global oxygen isotope record of benthic (deep ocean dwelling) foraminifera compiled by Lisieki and Raymo (2005), repeated in Figure 10a of Hansen *et al* (2007) for comparison with solar insolation changes over the same period, reveals long-term cooling and sea level fall, with superposed oscillations at a dominant frequency of 41 ky. The long-term cooling presumably is due, at least in part, to drawdown of atmospheric $CO_2$ by weathering that accompanied and followed the rapid growth of the Andes (Ghosh *et al* 2006), which was most rapid in the late Miocene. Changes in meridional heat transport may have contributed to the climate trend (Rind and Chandler 1991), but the $CO_2$ amount providing a global positive forcing seems unlikely to have been more than approximately 350-450 ppm (Dowsett *et al* 1994; Raymo *et al* 1996; Crowley 1996). Global mean temperature three million years ago was only 2-3°C warmer than today (Crowley 1996; Dowsett *et al* 1996), while sea level was 25 ± 10 m higher (Wardlaw and Quinn 1991; Barrett *et al* 1992; Dowsett *et al* 1994).

The Plio-Pleistocene record compiled by Lisieki and Raymo (2005) is fascinating to paleoclimatolgists as it clearly shows the expected dominance of global climate variations with the 41 ky cyclic variation of the tilt of the Earth's spin axis, increased tilt melting ice at both poles. When the planetary cooling reached a degree that allowed a large mid-latitude Northern Hemisphere (Laurentide) ice sheet, the periodicity necessarily became more complex, because of the absence of land area for a similar ice sheet in the Southern Hemisphere (Hansen *et al* 2007). However, the information of practical importance from the Plio-Pleistocene record is the implication of dramatic global climate change with only moderate global climate forcing. With global warming of only 2-3°C and $CO_2$ of perhaps 350-450 ppm it was a dramatically different planet, without Arctic sea ice in the warm seasons and sea level 25 ± 10 m higher.

Assuming a nominal 'Charney' climate sensitivity of 3°C equilibrium global warming for doubled $CO_2$, BAU scenarios yield a global warming at least of the order of 3°C by the end of this century. However, the Charney sensitivity is the equilibrium (long-term) global response when only fast feedback processes (changes of sea ice, clouds, water vapor and aerosols in response to climate change) are included (Hansen *et al* 2007). Actual global warming would be larger as slow feedbacks come into play. Slow feedbacks include increased vegetation at high latitudes, ice sheet shrinkage, and terrestrial and marine greenhouse gas emissions in response to global warming.

In assessing likely effects of warming of 3°C, it is useful to note the effects of the 0.7°C warming in the past century (Hansen *et al* 2006a). This warming already produces large areas of summer melt on Greenland and significant melt on West Antarctica. Global warming of several more degrees, with its polar amplification, would have both Greenland and West Antarctica bathed in summer melt for extended melt seasons.

The IPCC (2007) midrange projection for sea level rise this century is 20-43 cm [8-17 inches] and its full range is 18-59 cm [7-23 inches]. IPCC notes that they are unable to evaluate possible dynamical responses of the ice sheets, and thus do not include any possible "rapid dynamical changes in ice flow". Yet the provision of such specific numbers for sea level rise encourages a predictable public response that projected sea level change is moderate, and indeed smaller than in IPCC (2001). Indeed, there have been numerous media reports of "reduced" sea level rise predictions, and commentators have denigrated suggestions that business-as-usual greenhouse gas emissions may cause sea level rise measured in meters.

However, if these IPCC numbers are taken as predictions of actual sea level rise, as they have been by the public, they imply that the ice sheets can miraculously survive a BAU climate forcing assault for a period of the order of a millennium or longer. This is not entirely a figment of the IPCC decision to provide specific numbers for only a portion of the problem, while demurring from any quantitative statement about the most important (dynamical) portion of the problem. Undoubtedly there are glaciologists who anticipate such long response times, because their existing ice sheet models have been designed to match paleoclimate changes, which occur on millennial time scales.

However, Hansen *et al* (2007) show that the typical ~6ky time scale for paleoclimate ice sheet disintegration reflects the half-width of the shortest of the weak orbital forcings that drive the climate change, not an inherent time scale of ice sheets for disintegration. Indeed, the paleoclimate record contains numerous examples of ice sheets yielding sea level rise of several meters per century, with forcings smaller than that of the BAU scenario. The problem with the paleoclimate ice sheet models is that they do not generally contain the physics of ice streams, effects of surface melt descending through crevasses and lubricating basal flow, or realistic interactions with the ocean.

Rahmstorf (2007) has noted that if one uses observed sea level rise of the past century to calibrate a linear projection of future sea level, BAU warming will lead to sea level rise of the order of one meter in the present century. This is a useful observation, as it indicates that sea level change would be substantial even without non-linear collapse of an ice sheet. However, this approach cannot be taken as a realistic way of projecting likely sea level rise under BAU forcing. The linear approximation fits the past sea level change well for the past century only because the two terms contributing significantly to sea level rise were (1) thermal expansion of ocean water and (2) melting of alpine glaciers.

Under BAU forcing in the 21$^{st}$ century, sea level rise undoubtedly will be dominated by a third term (3) ice sheet disintegration. This third term was small until the past few years, but it is has at least doubled in the past decade and is now close to 1 mm/year, based on gravity satellite measurements discussed above. As a quantitative example, let us say that the ice sheet contribution is 1 cm for the decade 2005-2015 and that it doubles each decade until the West Antarctic ice sheet is largely depleted. That time constant yields sea level rise of the order of 5 m this century. Of course I can not prove that my choice of a 10 year doubling time for non-linear response is accurate, but I am confident that it provides a far better estimate than a linear response for the ice sheet component of sea level rise.

An important point is that the non-linear response could easily run out of control, because of positive feedbacks and system inertias. Ocean warming and thus melting of ice shelves will continue after growth of the forcing stops, because the ocean response time is long and the temperature at depth is far from equilibrium for current forcing. Ice sheets also have inertia and are far from equilibrium: and as ice sheets disintegrate their surface moves lower, where it is warmer, subjecting the ice to additional melt. There is also inertia in energy systems: even if it is decided that changes must be made, it may require decades to replace infrastructure.

The nonlinearity of the ice sheet problem makes it impossible to accurately predict sea level change on a specific date. However, as a physicist, I find it almost inconceivable that BAU climate change would not yield a sea level change measured in meters on the century time scale. The threat of large sea level change is a principal element in our argument (Hansen *et al* 2006a,b, 2007) that the global community must aim to keep additional global warming less than 1°C above 2000 temperature. In turn, this implies a $CO_2$ limit of about 450 ppm, or less. Such scenarios are dramatically different than BAU, requiring almost immediate changes to get on a fundamentally different energy and greenhouse gas emissions path.

**5. Reticence**

Is my perspective on this problem really so different than that of other members of the relevant scientific community? Based on interactions with others, I conclude that there is not such a great gap between my position and that of most, or at least much, of the relevant community. The apparent difference may be partly a natural reticence to speak out, which I attempt to illuminate via specific examples.

In the late 1980s Richard Kerr wrote an article titled "Hansen vs. the World on the Greenhouse Threat", reporting on a scientific conference in Amherst, Massachusetts. One may have surmised strong disagreement with my assertion (to Congress) that the world had entered a period of strong warming due to human-made greenhouse gases. But participants told Kerr "if there were a secret ballot at this meeting on the question, most people would say the greenhouse warming is probably there." And "what bothers us is that we have a scientist telling congress things that we are reluctant to say ourselves."

That article made me notice right away a difference between scientists and 'normal people'. A non-scientist friend from my hometown, who had congratulated me after my congressional testimony, felt bad after he saw the article by Kerr. He obviously believed that I had been shown to be wrong. However, I thought Kerr did a good job of describing the various perspectives, and made it clear, at least between the lines, that differences were as much about reticence to speak as about scientific interpretations.

IPCC reports may contain a reticence in the sense of being extremely careful about making attributions. This characteristic is appropriately recognized as an asset that makes IPCC conclusions authoritative and widely accepted. It is probably a necessary characteristic, given that the IPCC document is produced as a consensus among most nations in the world and represents the views of thousands of scientists.

Kerr (2007) describes a specific relevant example, whether IPCC should include estimates of dynamical ice sheet loss in their projections: "too poorly understood, IPCC authors said", and "overly cautious – (dynamical effects) could raise sea level much faster than IPCC was predicting" some scientists responded. Kerr goes on to say "almost immediately, new findings have emerged to support IPCC's conservative position." Glaciologist Richard Alley, an IPCC lead author, said "Lots of people were saying we [IPCC authors] should extrapolate into the future, but we dug our heels in at the IPCC and said that we don't know enough to give an answer."

**6. Our Legacy**

Reticence is fine for IPCC. And individual scientists can choose to stay within a comfort zone, not needing to worry that they say something that proves to be slightly wrong. But perhaps we should also consider our legacy from a broader perspective. Do we not know enough to say more?

Confidence in a scientific inference can be built from many factors. For climate change these include knowledge gained from studying paleoclimate changes, analysis of how the Earth has responded to forcings on various time scales, climate simulations and tests of these against observations, detailed study of climate change in recent decades and how the nature of observed change compares with expectations, measurements of changes in atmospheric composition and calculation of implied climate forcings, analysis of ways in which climate response varies among different forcings, quantitative data on different feedback processes and how these compare with expectations, and so on.

Can the broader perspective drawn from various sources of information allow us to 'see the forest for the trees', to 'separate the wheat from the chaff'? That a glacier on Greenland slowed after speeding up, used as 'proof' that reticence is appropriate, is little different than the common misconception that a cold weather snap disproves global warming. Spatial and temporal fluctuations are normal, short-term expectations for Greenland glaciers are different from long-term expectations for West Antarctica. Integration via the gravity satellite measurements puts individual glacier fluctuations in proper perspective. The broader picture gives strong indication that ice sheets will, and are already beginning to, respond in a nonlinear fashion to global warming. There is enough information now, in my opinion, to make it a near certainty that IPCC BAU climate forcing scenarios would lead to disastrous multi-meter sea level rise on the century time scale.

There is, in my opinion, a huge gap between what is understood about human-made global warming and its consequences, and what is known by the people who most need to know, the public and policy makers. IPCC is doing a commendable job, but we need something more. Given the reticence that IPCC necessarily exhibits, there need to be supplementary mechanisms. The onus, it seems to me, falls on us scientists as a community.

Important decisions are being made now and in the near future. An example is the large number of new efforts to make liquid fuels from coal, and a resurgence of plans for energy intensive "cooking" of tar-shale mountains to squeeze out liquid hydrocarbon fuels. These are just the sort of actions needed to preserve a BAU greenhouse gas path indefinitely. We know enough about the carbon cycle to say that at least of the order of a quarter of the $CO_2$ emitted in burning fossil fuels under a BAU scenario will stay in the air "forever", the latter defined practically as more than 500 years. Readily available conventional oil and gas are enough to take atmospheric $CO_2$ to a level of the order of 450 ppm.

In this circumstance it seems vital that we provide the best information we can about the threat to the great ice sheets posed by human-made climate change. This information, and necessary caveats, should be provided publicly, and in plain language. The best suggestion I can think of is for the National Academy of Sciences to carry out a study, in the tradition of the Charney and Cicerone reports on global warming. I would be glad to hear alternative suggestions.

**Acknowledgments**

I thank Tad Anderson, Mark Bowen, Svend Brandt-Erichsen, Jost Heintzenberg, John Holdren, Ines Horovitz, Bruce Johansen, Ralph Keeling, John Lyman, Maureen Raymo, Christopher Shuman, Richard Somerville, and Bob Thomas for comments on a draft version of this paper.